\documentclass[lettersize,journal]{IEEEtran}

\PassOptionsToPackage{hyphens}{url}
\usepackage{setspace}
\usepackage{tabularx}
\usepackage{amsmath}
\usepackage{amsthm}
\usepackage{amssymb}    

\usepackage{makecell}
\usepackage{longtable}
\usepackage{booktabs}
\usepackage{float}
\usepackage[table]{xcolor}   
\usepackage{multirow}
\usepackage{stfloats}

\usepackage[ruled,vlined]{algorithm2e}  

\usepackage{graphicx}
\usepackage{subfig}
\usepackage{url}
\usepackage{cite}
\usepackage{textcomp}

\usepackage{flushend}
\usepackage[colorlinks,linkcolor=red,anchorcolor=green,citecolor=blue]{hyperref}
\usepackage{cleveref}

\usepackage{bbding}
\usepackage{verbatim}
\crefname{figure}{fig}{figures}
\Crefname{figure}{Fig}{Figures}
\begin{document}
\title{SkyChain Intelligence: A Blockchain-Secured Multi-Agent DRL Framework for Low-Altitude Embodied Artificial Intelligence}

\author{Haoxiang Luo, Tianqi Jiang, Ruichen Zhang, Yinqiu Liu, Gang Sun,~\IEEEmembership{Senior Member,~IEEE},\\Hongfang Yu,~\IEEEmembership{Senior Member,~IEEE}, Abbas Jamalipour,~\IEEEmembership{Fellow,~IEEE}, and Dong In Kim,~\IEEEmembership{Life Fellow,~IEEE}
\thanks{H. Luo is with the WeBank-NTU Joint Research Institute on Fintech, Nanyang Technological University, Singapore 639798, and also with the College of Computing and Data Science, Nanyang Technological University, Singapore 639798 (e-mail:haoxiang.luo@ntu.edu.sg). T. Jiang is with the School of Science and Engineering,
The Chinese University of Hong Kong, Shenzhen 518172, China (e-mail: tikyttqq@gmail.com).  R. Zhang and Y. Liu are with the College of Computing and Data Science, Nanyang Technological University, Singapore 639798 (e-mail: ruichen.zhang@ntu.edu.sg; yinqiu001@e.ntu.edu.sg.  G. Sun  (Corresponding Author) and H. Yu are with the School of Information and Communication Engineering, University of Electronic Science and Technology of China, Chengdu 611731, China (e-mail: \{gangsun, yuhf\}@uestc.edu.cn). A. Jamalipour is with the School of Electrical and Computer Engineering, University of Sydney, Australia, and with the Graduate School of Information Sciences, Tohoku University, Japan (e-mail: a.jamalipour@ieee.org). D. I. Kim is with the Department of Electrical and Computer Engineering, Sungkyunkwan University, Suwon 16419, South Korea (e-mail: dongin@skku.edu).}

}



\maketitle

\begin{abstract}
With the rapid development of the Low-Altitude Economy (LAE) ecosystem, Low-Altitude Embodied Artificial Intelligence (LAEAI) agents have become the core carriers of autonomous aerial services, thereby enabling dynamic Low-altitude Computility Networks (LACNets) for distributed computing resource sharing. However, resource-constrained LAEAI agents in decentralized LACNets face a fundamental trilemma of autonomy, security, and efficiency. Existing solutions primarily focus on either optimizing computational performance or enhancing security in isolation, failing to address the inherent trade-offs among trust, performance, and overhead in untrusted dynamic environments with malicious agents. To tackle this challenge, this paper proposes SkyChain Intelligence, a holistic framework that synergistically integrates agentic AI, consortium blockchain, and Multi-Agent Deep Reinforcement Learning (MADRL). We design a lightweight blockchain-based decentralized trust management system with a dynamic reputation mechanism and develop a hybrid-action-space MADDPG algorithm that embeds on-chain reputation scores into the reward function to jointly optimize offloading decisions, resource allocation, and drone 3D trajectories. Extensive simulations demonstrate that our framework outperforms state-of-the-art baselines in task completion latency and energy consumption, while achieving a $94.1\%$ task completion rate in the baseline scenario and stable convergence within $300$ training episodes. This work provides a viable path for building secure, autonomous, and efficient machine-to-machine computing ecosystems in the low-altitude domain.

\end{abstract}

\begin{IEEEkeywords}
Agentic AI, Low-Altitude Embodied Artificial Intelligence (LAEAI), Low-Altitude Computility Networks (LACNets), Multi-Agent Deep Reinforcement Learning (MADRL), blockchain.
\end{IEEEkeywords}

\section{Introduction} \label{sec-I}
\subsection{Background}
\IEEEPARstart {T}{he} world is on the cusp of a transformative shift driven by the Low-Altitude Economy (LAE), an emerging economic paradigm centered around various flight activities of civil manned and unmanned aircraft in the airspace typically below 1,000 meters \cite{li2024unauthorized}. This nascent domain is poised to unlock immense economic and social value by introducing unprecedented efficiency and capabilities across numerous sectors, including logistics and delivery, precision agriculture, urban air mobility, emergency response, and environmental monitoring \cite{cai2025large}.

At the heart of this revolution is the advent of Low-Altitude Embodied Artificial Intelligence (LAEAI) \cite{yang2024embodied}. LAEAI refers to autonomous physical agents, such as Unmanned Aerial Vehicles (UAVs) and electric Vertical Take-Off and Landing (eVTOL) aircraft, that can perceive, reason, plan, and act within the physical world. Unlike traditional passive AI, LAEAI does not merely follow preset rules or react to stimuli. It exhibits agentic capabilities, pursuing goals in an autonomous, proactive, and adaptive manner, capable of making independent decisions and executing tasks in dynamic and unstructured environments \cite{zhang2024generative}.

\begin{figure*}[!t]
   \centering
   \includegraphics[width=5 in]{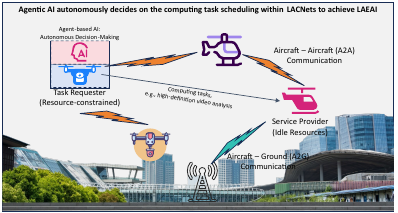}
   \caption{LAEAI. Each unmanned aircraft will be equipped with agentic AI, which will autonomously determine the computing tasks scheduling in the LACNets to achieve the goals of LAEAI.}
   \label{fig0}
    \vspace{-0.4cm}
\end{figure*}

Despite the immense potential of LAEAI agents, they are resource-constrained. A single aircraft faces stringent limitations in terms of computational power, onboard energy, sensor range, and communication bandwidth \cite{xue2025joint}. To execute increasingly complex tasks, such as real-time high-definition video analysis, collaborative environmental mapping, or distributed data processing, individual LAEAI agents must collaborate beyond the confines of their own capabilities \cite{luo2025toward}.
This necessity gives rise to the concept of Low-altitude Computility Networks (LACNets) \cite{luo2025real}. We define a LACNet as a dynamic, aircraft-assisted Mobile Edge Computing (MEC) network architecture that provides ``computility (computing utility)" on demand. As shown in Fig. \ref{fig0}, in this paradigm, a swarm of aircraft forms a self-organizing aerial computing network where some aircraft can offload their computation-intensive tasks to neighboring aircraft with idle resources \cite{wang2025multi}. This creates a flexible and resilient computing infrastructure that does not rely on ground-based stations, which is particularly crucial in remote areas or post-disaster scenarios \cite{chen2022multi}.

\subsection{Research Motivation}

While the vision of LACNets is promising, its realization is confronted by a fundamental trilemma of autonomy, security, and efficiency.

First, LAEAI agents must make complex, decentralized decisions in real-time without constant human supervision \cite{wang2025cluster}. This requires them to possess true agentic capabilities, the ability to proactively plan, reason, and act to achieve high-level goals, not just execute pre-programmed instructions \cite{liu2025lameta}. Multi-Agent Deep Reinforcement Learning (MADRL) is a powerful tool for achieving such autonomous decision-making \cite{sun2025edge}, but it requires a well-defined objective function to guide the learning process.

Additionally, in a decentralized network composed of multiple agents, potentially belonging to different operators with competing or selfish behaviors, trust is the cornerstone of collaboration. How can one agent trust another to faithfully execute a computational task, not steal sensitive data, or fairly report its resource usage? Traditional centralized authorities are not only antithetical to the decentralized nature of UAV swarms but also represent a single point of failure \cite{luo2025convergence}. A decentralized trust management mechanism is therefore imperative \cite{luo2025weighted}.

Moreover, resource management in a LACNet is an exceedingly complex optimization problem. Jointly optimizing computational resources (CPU cycles), communication resources (spectrum, power), energy (battery life), and agent behavior (trajectory planning, task offloading decisions) constitutes a high-dimensional, non-convex, mixed-integer nonlinear programming problem. It is particularly intractable in a dynamically changing environment \cite{quirynen2024real}.

These three challenges are interconnected. An efficient system without security guarantees can have its autonomy exploited by malicious actors. A secure system that is inefficient will fail to meet the real-time demands of LAEAI applications. Therefore, any viable solution must address all three issues concurrently.

\subsection{Our Contributions}

To address the trilemma above, this paper proposes a holistic framework named SkyChain Intelligence, which synergistically integrates agentic AI, blockchain, and MADRL. Our framework is built upon a symbiotic relationship where each technology compensates for the inherent limitations of the others in this specific application context. The main contributions of this paper can be summarized as follows:

\begin{itemize}
    \item \textbf{Integrated Framework for Agentic LAEAI:} We propose the SkyChain Intelligence framework, which for the first time synergistically integrates agentic AI, blockchain, and MADRL. This provides a holistic solution that empowers LAEAI agents with the autonomous, goal-driven, and adaptive capabilities needed for secure and efficient computation offloading in LACNets.
     \item \textbf{Blockchain-based Decentralized Trust Management:} We design a trust management system built on a lightweight consortium chain. It features a dynamic reputation scoring mechanism managed by smart contracts and an efficient consensus, enabling reliable cooperation among LAEAI agents without a central authority.
     \item \textbf{Hybrid-Action MADRL for Complex Decision-Making:} 
     We develop a customized MADRL algorithm with a dual-head actor-critic architecture tailored for hybrid action spaces. It enables end-to-end joint optimization of discrete task offloading decisions, continuous resource allocation, and UAV 3D trajectory control in a fully decentralized manner, while embedding blockchain-derived reputation states into the policy learning loop to align agent decision-making with both system performance and trustworthiness.
     \item \textbf{Synergistic Trust-Performance Reward Mechanism:} We introduce a reward function that directly integrates blockchain-derived reputation scores into the MADRL learning process. This creates a powerful feedback loop. It incentivizes agents to learn policies that not only optimize for performance metrics but also actively build and maintain trust within the network.
\end{itemize}

\subsection{Paper Structure}

The remainder of this paper is organized as follows. Section \ref{sec-II} reviews related work. Section \ref{sec-III} establishes the detailed system model and formulates the problem mathematically. Section \ref{sec-Iv} elaborates on the design of the SkyChain Intelligence framework. Section \ref{sec-v} provides a security analysis of the proposed solution. Section \ref{sec-vi} presents and analyzes the simulation results in depth. Finally, Section \ref{sec-vii} concludes the paper and discusses future research directions.

\section{Related Works}\label{sec-II}

This section provides a critical review of the existing literature to identify the research gap that our work aims to fill. We will explore four key areas: resource management in aircraft-assisted MEC, DRL for aircraft networks, blockchain for aircraft and IoT security, and Federated Learning (FL) as a comparative approach. Table \ref{tab:laec_works} presents a comparison of the related work in these fields. 

\begin{table*}[!t]
\centering
\caption{Comparison of Related Works}
\label{tab:laec_works}

\renewcommand{\arrayrulewidth}{0.8pt} 
\renewcommand{\tabcolsep}{6pt} 

{\fontsize{8}{10}\selectfont 
\begin{tabular}{m{1.38cm}<{\centering}||m{2.56cm}<{\centering}|m{3.98cm}<{\centering}|m{1.2cm}<{\centering}|m{0.84cm}<{\centering}|m{0.99cm}<{\centering}|m{0.9cm}<{\centering}|m{1.16cm}<{\centering}|m{1cm}<{\centering}}
    \hline
    \hline
    \textbf{Related Work} & \textbf{Scenario} & \textbf{Problem} & \textbf{Blockchain} & \textbf{Agentic AI} & \textbf{Hybrid Action} & \textbf{MADRL} & \textbf{3D Trajectory} & \textbf{Resource Alloc.} \\
    \hline
    
    \cite{xue2025joint} & 3-layer heterogeneous LAENet & Delay and energy optimization under unstable links & \XSolidBrush & \XSolidBrush & \Checkmark & \XSolidBrush & \XSolidBrush & \Checkmark \\
    \hline
    \cite{luo2025real} & LACNets urban logistics & Trust and resource issues in aerial computility sharing & \Checkmark & \XSolidBrush & \XSolidBrush & \XSolidBrush & \XSolidBrush & \Checkmark \\
    \hline
    \cite{wang2025multi} & Multi-UAV relay and edge networks & Minimize MEC task delay in 3D environments & \XSolidBrush & \XSolidBrush & \Checkmark & \Checkmark & \Checkmark & \Checkmark \\
    \hline
       \cite{kang20203d} & UAV relaying in urban environments & Max-min throughput among users & \XSolidBrush & \XSolidBrush & \XSolidBrush & \XSolidBrush & \XSolidBrush & \Checkmark \\
    \hline
      \cite{fu2023federated} & UAV-assisted FL in remote environments & Minimize FL completion time and mitigate stragglers & \XSolidBrush & \XSolidBrush & \XSolidBrush & \XSolidBrush & \Checkmark & \Checkmark \\
    \hline
    
    \cite{chen2025maddpg} & UAV-assisted resource slicing & Joint user association/slicing with interference & \XSolidBrush & \XSolidBrush & \XSolidBrush & \Checkmark & \XSolidBrush & \Checkmark \\
    \hline
    \cite{wang2025joint} & UAV-assisted energy-sensitive tasks& Energy consumption/system complexity in scheduling & \XSolidBrush & \XSolidBrush & \XSolidBrush & \Checkmark & \Checkmark & \Checkmark \\
    \hline
  
    \cite{tummala2024efficient} & UAV-assisted edge for IoT & Balance energy/delay in multi-device offloading & \XSolidBrush & \XSolidBrush & \XSolidBrush & \XSolidBrush & \XSolidBrush & \Checkmark \\
    \hline
    \cite{chen2024multi} & UAV-assisted satellite edge computing & Cost minimization under coverage and resource constraints & \XSolidBrush & \XSolidBrush & \XSolidBrush & \XSolidBrush & \XSolidBrush & \Checkmark \\
    \hline

    \cite{jia2026blockchain} & Zero-Trust with multi-UAV clusters & Routing security and stability in dynamic UAV networks & \Checkmark & \XSolidBrush & \XSolidBrush & \Checkmark & \XSolidBrush & \Checkmark \\
    \hline
    
    SkyChain Intelligence (Ours) & LACNets with LAEAI swarms & Autonomy, security, and efficiency trilemma in MEC offloading & \Checkmark & \Checkmark & \Checkmark & \Checkmark & \Checkmark & \Checkmark \\
    \hline
    \hline
\end{tabular}}
 \vspace{-0.4cm}
\end{table*}

\subsection{Resource Management in Aircraft-Assisted MEC}

Aircraft-assisted MEC has been widely studied as an effective paradigm for providing flexible computing services to ground users \cite{wang2025multi}. Early works primarily focused on using traditional optimization methods to solve resource allocation problems. For instance, some studies employed techniques like convex optimization and block coordinate descent (BCD) to jointly optimize UAV trajectory, power allocation, and task partitioning to minimize latency or energy consumption \cite{kang20203d}. Furthermore, game theory has been widely applied to model the competitive or cooperative relationships among aircraft or users \cite{mkiramweni2019survey}, seeking optimal resource allocation strategies under Nash Equilibrium (NE) or Stackelberg Equilibrium (SE). These traditional methods perform well in static scenarios where the system model is relatively simple, and the environmental state is precisely known.

The primary limitation of these traditional optimization techniques is their inability to adapt to the highly dynamic and uncertain nature of LACNets \cite{cai2025large}. They often rely on complete and accurate global system information, which is impractical to obtain in a decentralized, mobile network of autonomous agents \cite{zheng2026agentvne}. Moreover, the joint optimization problems are frequently NP-hard, making real-time solutions computationally intractable \cite{xue2025joint}. This inflexibility and high computational overhead render them unsuitable for enabling the real-time, autonomous decision-making required by LAEAI agents, thereby motivating our adoption of a more adaptive, learning-based DRL approach

\subsection{Deep Reinforcement Learning for LAENets}

To overcome the limitations of traditional optimization methods, DRL has emerged as a powerful tool for solving dynamic decision-making problems in LAENets \cite{cai2025secure}. Researchers have applied various DRL algorithms, including value-based Deep Q-Networks (DQN) and policy-based algorithms like DDPG and Soft Actor-Critic (SAC), to address problems such as aircraft trajectory optimization, computation offloading, and resource allocation \cite{frattolillo2023scalable}, \cite{jia2026blockchain}. Given the distributed nature of LAENets, MADRL methods, such as MADDPG and MAPPO, have received more attention \cite{zheng2024safe}. Also, the classic MADDPG is incompatible with discrete-continuous hybrid action spaces. These works have successfully demonstrated the great potential of DRL in achieving autonomous decision-making for aircraft and improving system efficiency.

However, a significant and critical flaw in the existing body of DRL-based research is the near-universal assumption of a fully cooperative and trustworthy environment. These models do not account for the possibility of selfish or malicious nodes that may drop tasks, return false results, or act unreliably to conserve their resources \cite{lin2023drl}. In an open and economically-driven LAE, where LAEAI agents may belong to different, competing operators, this assumption is unrealistic and dangerous \cite{luo2025convergence}. This critical omission of a trust and security mechanism makes existing DRL solutions vulnerable and incomplete. It directly motivates to integration of a verifiable, decentralized trust layer into the DRL framework.

\subsection{Blockchain for Aircraft and IoT Security}
In contrast to DRL, which focuses on efficiency and autonomy, another line of research utilizes blockchain technology to enhance the security of aircraft and Internet of Things (IoT) systems. Numerous studies have explored how to use blockchain to ensure the security of UAV communications, data integrity, reliability of access control, and traceability of transactions \cite{luo2025real}. Given the resource constraints of edge devices like aircraft, research often focuses on designing lightweight consensus protocols, such as Practical Byzantine Fault Tolerance (PBFT) and Raft \cite{luo2024symbiotic} to replace the computationally intensive Proof-of-Work (PoW) \cite{luo2025wireless}. These works have successfully provided a security foundation for decentralized systems by creating immutable and transparent ledgers for transactions and data management \cite{luo2025weighted}.  

While blockchain provides a robust foundation for security and trust, existing research typically treats it as an isolated component. These studies often focus on static data management and security primitives, failing to integrate the security layer with the dynamic, real-time decision-making processes required for resource management \cite{luo2025wireless}. The inherent latency and energy overhead of blockchain operations are frequently ignored or oversimplified, yet they have a direct impact on the performance of time-sensitive LAEAI applications \cite{liu2024blockchain}. This disconnect between security and performance optimization is a major gap. Our framework addresses this by not only using blockchain for trust but also by making the AI agents aware of the blockchain's overhead, allowing them to learn intelligent policies that balance security needs with efficiency constraints. 

\subsection{Federated Learning in LAENets}

FL is another decentralized AI paradigm applied in LAENets \cite{fu2023federated}. It allows multiple devices, such as UAVs, to collaboratively train a shared machine learning model without sharing their raw data, thus protecting data privacy and reducing communication overhead \cite{che2024leveraging}. This is particularly useful when data is sensitive or when communication bandwidth is limited, both of which are common constraints in LAENets. The primary goal is to leverage distributed data to build a more accurate and robust global model than any single agent could train on its own \cite{che2023multimodal}. 

Although both FL and MADRL are decentralized AI technologies, they are designed to solve fundamentally different problems. The core objective of FL is collaborative model training, whereas the core objective of MADRL is collaborative decision control \cite{yuan2025digital}. In our LACNet scenario, the challenge is not to train a shared predictive model, but to enable each LAEAI agent to learn its optimal policy for making sequential decisions about task offloading, resource allocation, and trajectory control \cite{cai2025large}, \cite{zhang2024generative}. Therefore, this crucial distinction justifies our choice of MADRL as the appropriate framework for empowering LAEAI agents with autonomous decision-making capabilities.

\section{System Model and Problem}\label{sec-III}

\begin{figure}[!t]
\centering
 \includegraphics[width=3.5 in]{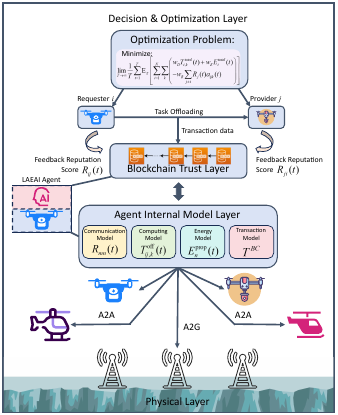}
   \caption{SkyChain Intelligence framework. It consists of the physical layer for aircraft, the agent layer within the aircraft, the blockchain trust layer, and the final decision and optimization layer.}
\label{fig1}
 \vspace{-0.4cm}
\end{figure}

This section establishes the rigorous mathematical foundation for our SkyChain Intelligence framework, as shown in Fig. \ref{fig1}, covering the network architecture, communication, computation, energy consumption, and blockchain-based trust models. A summary of key notations is provided in Table \ref{tab:notations}.

\begin{table}[t!]
\centering
\caption{Summary of Notations}
\label{tab:notations}
    \renewcommand{\arrayrulewidth}{0.5pt}
    \renewcommand{\tabcolsep}{10pt}
    
    {\fontsize{8}{10}\selectfont
    \begin{tabular}{m{1.95cm}<{\centering}||m{5.515cm}}
        \hline\hline
\textbf{Symbol} & \textbf{Definition/Physical Meaning} \\ \hline
$a_{ijk}(t)$ & Binary offloading decision variable \\
$B$ & Communication channel bandwidth \\
$C_k$ & Computation density of task $k$ \\
$D_k$ & Data size of task $k$ \\
$E^{BC}, T^{BC}$ & Energy and latency for a blockchain transaction \\
$E^{comp}, E^{trans}$ & Energy for computation and communication \\
$E^{prop}, P^{prop}$ & Energy and power for aircraft propulsion \\
$f_i^{loc}, f_j^{edge}$ & CPU frequency for local and edge computation \\
$H_n$ & Flight altitude of aircraft $n$ \\
$\mathcal{M}, M$ & Set and number of ground task generators \\
$\mathcal{N}, N$ & Set and number of LAEAI agents (aircraft) \\
$N_0$ & Noise power spectral density \\
$o_n(t), a_n(t), r_n(t)$ & Observation, action, and reward for agent $n$ \\
$P_{LoS}(\cdot)$ & Probability of a LoS link \\
$P_m, P_n^{trans}$ & Transmit power of device $m$ and aircraft $n$ \\
$PL_{nm}(t)$ & Path loss between aircraft $n$ and device $m$ \\
$Q_{\phi_n}$ & Actor and Critic networks for agent $n$ \\
$R_{mn}(t)$ & Data transmission rate from $m$ to $n$ \\
$\mathcal{R}_n(t)$ & Reputation score of aircraft $n$ at time $t$ \\
$t, T$ & Time slot index and total number of time slots \\
$T_{i,k}^{loc}, T_{ij,k}^{off}$ & Latency for local and offloaded computation \\
$T_k^{\max}$ & Max latency of task $k$ \\
$w_D, w_E, w_R$ & Weighting factors for delay, energy, and reputation \\
$\mathbf{w}_m$ & Fixed 3D position of ground device $m$ \\
$\mathbf{q}_n(t)$ & 3D position of aircraft $n$ at time $t$ \\
$\theta_{nm}(t)$ & Elevation angle between aircraft $n$ and device $m$ \\
$\mu_{\theta_n}$ & Actor network for agent $n$ \\ \hline\hline
    \end{tabular}}
     \vspace{-0.4cm}
\end{table}

\subsection{Network Structure}

We consider a Low-altitude Computility Network (LACNet) consisting of a set of $N$ LAEAI agents (aircraft), denoted as $\mathcal{N} = \{1,..., N\}$. These UAVs operate in a three-dimensional Cartesian space over a finite time horizon $T$, which is discretized into time slots $t \in \{1,..., T\}$. The position of aircraft $n$ at time slot $t$ is denoted by $\mathbf{q}_n(t) = [x_n(t), y_n(t), H_n]$, where $H_n$ is its quasi-static flight altitude. The network also includes a set of $M$ ground task generators, e.g., IoT devices or ground UAV stations, denoted as $\mathcal{M} = \{1,..., M\}$, with fixed positions $\mathbf{w}_m$. In this system, aircraft can act as both requesters of computation tasks and providers of computation services, forming a peer-to-peer computation market.

\subsection{Air-to-Ground (A2G) Communication Model}

In complex low-altitude environments such as cities, a deterministic Line-of-Sight (LoS) channel model is unrealistic \cite{wang2025multi}. Therefore, we adopt a more realistic probabilistic LoS channel model.
The probability of an LoS link between aircraft $n$ at position $\mathbf{q}_n(t)$, and ground device $m$ at position $\mathbf{w}_m$ depends on the elevation angle $\theta_{nm}(t)$ between them, namely,
\begin{equation}
\theta_{nm}(t) = \arctan\left(\frac{H_n}{\|\mathbf{q}_n^{xy}(t) - \mathbf{w}_m\|}\right),
\end{equation}
where $\mathbf{q}_n^{xy}(t)$ is the projection of aircraft $n$ onto the 2D plane.

According to the model in \cite{gapeyenko2021line}, the LoS probability can be expressed as a sigmoid function,
\begin{equation}
P_{LoS}(\theta_{nm}(t)) = \frac{1}{1 + a \exp(-b(\theta_{nm}(t) - a))},
\end{equation}
where $a$ and $b$ are environment-specific parameters, e.g., urban, suburban, that reflect the statistical properties of building density and height.

Additionally, the total path loss $PL_{nm}(t)$ is a weighted average of the LoS and Non-Line-of-Sight (NLoS) path losses \cite{wang2025multi}, that is,
\begin{equation}
\begin{aligned}
PL_{nm}(t) =&P_{LoS}(\theta_{nm}(t)) \cdot PL_{nm}^{\text{LoS}}(t) \\&+ (1 - P_{LoS}(\theta_{nm}(t))) \cdot PL_{nm}^{\text{NLoS}}(t)
\end{aligned}
\end{equation}
where the mean path losses (in dB) for LoS and NLoS links are given by
\begin{equation}
\begin{aligned}
\bar{PL}^{\text{LoS/NLoS}}_{nm}(t) = &20\log_{10}(d_{nm}(t)) + 20\log_{10}(f_c) \\& +20\log_{10}\left(\frac{4\pi}{c}\right) + \eta_{\text{LoS/NLoS}},\\
\end{aligned}
\end{equation}

\begin{equation}
d_{nm}(t) = \|\mathbf{q}_n(t) - \mathbf{w}_m\|,
\end{equation}
where $d_{nm}(t)$ is the distance between the aircraft and the ground device, $f_c$ is the carrier frequency, $c$ is the speed of light, and $\eta_{\text{LoS}}$ and $\eta_{\text{NLoS}}$ are additional attenuation factors for LoS and NLoS links, respectively.

Building upon this mean path loss formulation, we model long-term shadowing loss due to random building blockages as a zero-mean log-normal random variable:
\begin{equation}
PL^{\text{LoS/NLoS}}_{nm}(t)= \bar{PL}^{\text{LoS/NLoS}}_{nm}(t) + \xi_{\text{LoS/NLoS}},
\end{equation}
where $\xi_{\text{LoS/NLoS}} \sim \mathcal{N}(0, \sigma_{\text{LoS/NLoS}}^2)$ denotes shadowing fluctuation in the dB domain \cite{al2014optimal}. The terms $\eta_{\text{LoS}}$ and $\eta_{\text{NLoS}}$ in Eq. (4) represent the mean additional attenuation, including average shadowing and clutter loss, for each link state. The probabilistic LoS/NLoS switching in Eqs. (2)--(3) further captures long-term blocking state transitions driven by UAV mobility.

Moreover, according to the Shannon-Hartley theorem, the achievable data rate for offloading task $k$ from device $m$ to aircraft $n$ is
\begin{equation}
R_{mn}(t) = B \log_2\left(1 + \frac{P_m |h_{mn}(t)|^2}{N_0}\right),
\end{equation}
where $B$ is the channel bandwidth, $P_m$ is the transmission power of device $m$, $N_0$ is the noise power spectral density. 

We further decompose the instantaneous channel gain to characterize short-term small-scale fading:
\begin{equation}
|h_{mn}(t)|^2 = 10^{-PL_{nm}(t)/10} \cdot \alpha^2(t)
\end{equation}
where $|h_{mn}(t)|^2$ is the channel gain, $\alpha(t)$ is the small-scale fading envelope. We define the Rician K-factor as the power ratio of the LoS component to scattered multipath components $
K = \frac{P_{\text{LoS}}}{P_{\text{scatter}}}$.
For LoS paths, $\alpha(t)$ follows a Rician distribution with an elevation-dependent $K$; for NLoS paths, $\alpha(t)$ reduces to a Rayleigh distribution ($K=0$) due to rich multipath scattering \cite{gapeyenko2021line}. 

Moving beyond A2G links, Air-to-Air (A2A) channels between UAVs exhibit distinct propagation properties due to the elevated position of both transceivers. They are typically modeled as Rician fading channels with a much higher LoS probability \cite{salhab2021accurate}.
Also, we denote the Rician K-factors of A2G and A2A LoS links as $K_{\text{A2G}}$ and $K_{\text{A2A}}$, respectively. 
$K_{\text{A2G}}$ varies with elevation angle due to ground clutter, while $K_{\text{A2A}}$ remains stable as both transceivers operate above rooftop level with fewer obstacles \cite{li2022measurement}.

For high-mobility scenarios where UAVs move independently, the maximum Doppler shift induced by radial relative motion is formulated as 
$f_d^{\text{max}} = \frac{v_r f_c}{c}$, 
where $v_r$ m/s denotes the maximum radial relative velocity. The corresponding channel coherence time follows the standard approximation $T_c \approx \frac{9}{16\pi f_d^{\text{max}}}$, which is far longer than the slot duration $\Delta t$ (i.e., quasi-static fading model) for the considered system carrier frequency \cite{zeng2016wireless}. As a result, Doppler variations are fully captured by updating channel states slot-by-slot with real-time UAV positions.

\subsection{Computation Task and Offloading Model}

Each computation task $k$ initiated or received by aircraft $n$ is defined by a tuple $\mathcal{T}_k = \{D_k, C_k, T_k^{\max}\}$, representing the task's data size (in bits), computation density (CPU cycles/bit), and maximum tolerable latency (in seconds), respectively.

We define a discrete offloading decision variable $a_{ijk}(t) \in \{0, 1\}$, where $a_{ijk}(t)=1$ indicates that aircraft $i$ offloads task $k$ to aircraft $j$ for computation at time slot $t$, with $\sum_{j \in \mathcal{N}} a_{ijk}(t) = 1$. The case $j=i$ represents local computation.
Then, we set the local computation latency of aircraft $n$ for task $k$ to be
    \begin{equation}
    T_{i,k}^{\text{loc}}(t) = \frac{D_k C_k}{f_i^{\text{loc}}},
    \end{equation}
    where $f_i^{\text{loc}}$ is the CPU frequency of aircraft $i$ used for local computation.
The total computation latency for offloading to aircraft $j$ includes transmission latency and computation latency at aircraft $j$, namely,
    \begin{equation}
    T_{ij,k}^{\text{off}}(t) = T_{ij,k}^{\text{trans}}(t) + T_{j,k}^{\text{comp}}(t) = \frac{D_k}{R_{ij}(t)} + \frac{D_k C_k}{f_j^{\text{edge}}},
    \end{equation}
    where $R_{ij}(t)$ is the transmission rate from aircraft $i$ to $j$, and $f_j^{\text{edge}}$ is the CPU frequency allocated by aircraft $j$ to this task.

\subsection{Energy Consumption Model}
The total energy consumption of the system consists of computation, communication, flight, and blockchain interaction.

First, for the computation energy, the processing task $k$ is proportional to the square of the CPU frequency,
    \begin{equation}
    E_{n,k}^{\text{comp}} = \kappa (f_n)^2 D_k C_k
    \end{equation}
    where $\kappa$ is the effective capacitance coefficient related to the processor chip architecture.
    
Second, the communication energy for task $k$ is
    \begin{equation}
    E_{ij,k}^{\text{trans}} = P_i^{\text{trans}} \cdot T_{ij,k}^{\text{trans}}(t)
    \end{equation}
    where $P_i^{\text{trans}}$ is the transmission power of aircraft $i$.
    
Third, for a rotary-wing aircraft, the aircraft propulsion energy is a complex function of its flight speed $v_n(t)$ \cite{liu2024deep}, that is
    \begin{equation}
    \begin{aligned}
    P_n^{\text{prop}}(v_n(t)) = &P_0\left(1 + \frac{3v_n(t)^2}{U_{\text{tip}}^2}\right) \\&+ P_i\left(\sqrt{1 + \frac{v_n(t)^4}{4v_0^4}} - \frac{v_n(t)^2}{2v_0^2}\right)^{1/2} + \\&\frac{1}{2}d_0\rho s A v_n(t)^3,
    \end{aligned}
    \end{equation}
    where $P_0$ and $P_i$ are the blade profile power and induced power in hover, respectively, $U_{\text{tip}}$ is the blade tip speed, $v_0$ is the mean rotor induced velocity in hover, and others are aerodynamic constants. The propulsion energy within time slot $\Delta t$ is 
    
      \begin{equation}
      E_n^{\text{prop}}(t) = P_n^{\text{prop}}(v_n(t)) \cdot \Delta t.
       \end{equation}
    
Finally, we model blockchain transaction energy as the sum of fixed local validation energy and dynamic communication energy, to capture the impacts of wireless channel fading and UAV mobility.
Local validation energy $E_{\text{comp}}^{\text{BC}}$ (signature verification, smart contract execution) remains a constant.

We define $R_{\text{min}}$ as the minimum A2A rate for delivering one consensus message per slot. An outage occurs when $R_{n}(t) < R_{\text{min}}$ and triggers retransmissions. For A2A Rician channels, the outage probability has a closed-form expression~\cite{goldsmith2005wireless}:
\begin{equation}
P_{\text{out},n}(t) = 1 - Q_1\left( \sqrt{2K_{\text{A2A}}}, \sqrt{2(K_{\text{A2A}}+1) \cdot \frac{\gamma_{\text{th}}}{\gamma_n(t)}} \right),
\end{equation}
where $Q_1(\cdot,\cdot)$ is the first-order Marcum Q-function, $\gamma_{\text{th}} = 2^{R_{\text{min}}/B} - 1$ is the Signal-to-Noise Ratio (SNR) threshold, and $\gamma_n(t)$ is the received SNR at the time slot $t$ with the quasi-static fading model.

Incorporating expected retransmission overhead for wireless consensus, the per-transaction communication energy for node $n$ in a shard of size $N_s$ at slot $t$ is:
\begin{equation}
E_{\text{trans},n}^{\text{BC}}(t) = \frac{N_s \cdot P_n^{\text{trans}} \cdot \frac{S_{\text{msg}}}{R_{n}(t)}}{1 - P_{\text{out},n}(t)},
\end{equation}
where $S_{\text{msg}}$ is the consensus message size, and $R_{n}(t)$ is the achievable A2A rate from Section III-B. The denominator denotes the expected number of transmission attempts for successful delivery.
Thus, total blockchain energy can be expressed as $E_n^{\text{BC}}(t) = E_{\text{comp}}^{\text{BC}} + E_{\text{trans},n}^{\text{BC}}(t)$.

\subsection{Blockchain-based Trust and Transaction Model}

We adopt a consortium chain architecture, where all participating LAEAI agents are permissioned nodes. This provides better privacy, controllability, and performance than a public chain.
To minimize latency and energy overhead, we choose a lightweight consensus protocol, namely the sharded Symbiotic PBFT (SS-PBFT) \cite{luo2025convergence}. In this consensus, a symbiotic and reciprocal transmission relationship was established between the nodes, and the consensus efficiency was significantly improved and energy consumption was reduced through the form of sharding.

Additionally, a key smart contract is used to manage the reputation score $\mathcal{R}_n(t) \in [0, 1]$ of each aircraft $n$. After aircraft $j$ completes a computation task for aircraft $i$, the task requester $i$ submits feedback (e.g., 1 for success, 0 for failure) to the smart contract. We design a dual-track reputation mechanism, which is elaborated as follows:
\emph{1) Short-term reputation} $\mathcal{R}_j^{\text{s}}(t)$ adopts a large update factor $\beta_s\in (0, 1]$ to sensitively capture recent behavioral changes, enabling rapid identification of degraded or malicious nodes;
\emph{2) Long-term reputation} $\mathcal{R}_j^{\text{l}}(t)$ adopts a small update factor $\beta_l\in (0, 1]$ ($\beta_l < \beta_s$) to accumulate historical interaction records, anchoring reputation to long-term performance to resist intermittent misbehavior and whitewashing.

The update rules are:
\begin{equation}
\mathcal{R}_j^{\text{s}}(t) =
\begin{cases}
(1-\beta_s)\mathcal{R}_j^{\text{s}}(t-1) + \beta_s \cdot \text{Feedback}, & \text{If Normal}, \\
\mathcal{R}_j^{\text{s}}(t-1) - \beta_s \cdot \mathcal{R}_{\text{penalty}}, & \text{If Malicious}, \\
\mathcal{R}_{\text{init}}, & \text{New Node},
\end{cases}
\end{equation}
\begin{equation}
\mathcal{R}_j^{\text{l}}(t) = (1-\beta_l)\mathcal{R}_j^{\text{l}}(t-1) + \beta_l \cdot \text{Feedback},
\end{equation}
 where $\text{Feedback}\in \{0, 1\}$. $\mathcal{R}_{penalty} \in (0, 1]$  represents the penalty coefficient for malicious behavior. $\mathcal{R}_{init}$ is the initial reputation of the new node, which prevents the new node from immediately obtaining high trust privileges.
The final comprehensive reputation score is the weighted fusion of the two tracks:
\begin{equation}
\mathcal{R}_j(t) = \lambda \mathcal{R}_j^{\text{short}}(t) + (1-\lambda) \mathcal{R}_j^{\text{long}}(t)
\end{equation}
where $\lambda \in (0,1)$ balances sensitivity to immediate behavior and reliance on long-term credit.
    
Meanwhile, a computation offloading task is encapsulated as a transaction and recorded on the blockchain. The transaction content includes \texttt{\{RequesterID, ProviderID, TaskID, ResourcePrice, ComputationResultHash\}}. This provides an immutable audit trail for all interactions.

Then, the total latency for a blockchain transaction is
  \begin{equation}
T^{\text{BC}} = T^{\text{prop}} + T^{\text{consensus}},
\end{equation}
where $T^{\text{prop}}$ is the network propagation delay and $T^{\text{consensus}}$ is the time required to reach consensus. For the SS-PBFT, the consensus time is typically in the order of seconds due to the small scope of consensus, which is much lower than that of the original PBFT. To adapt to high UAV mobility and frequent topology changes, we add two mobility-aware optimizations to SS-PBFT:

\emph{1) Topology-aware dynamic sharding:} Shards are re-clustered per time slot via greedy clustering \cite{luo2025weighted}, where only node pairs with A2A LoS probability $P_{\text{LoS}}(t) \ge P_{\text{th}}^{\text{shard}}$ are grouped into the same shard to stabilize intra-shard links. The highest-reputation, lowest-mobility node is elected as shard primary to reduce coordinator failures.

\emph{2) Checkpoint-based fast synchronization:} Nodes generate compact state checkpoints every $M$ blocks. Reconnected nodes only sync incremental blocks after the latest checkpoint instead of the full ledger. For temporary partitions, shards run independent consensus, and ledgers are merged by reputation-weighted voting upon reconnection to ensure eventual consistency.

\subsection{Adversarial Model}
This part formally defines the adversary capabilities and types, and attack objectives of the system. It provides a unified theoretical framework for subsequent security analysis and algorithm design.
\subsubsection{Adversary Capabilities}
We define the adversary as attackers who possess the following capabilities:

\begin{itemize}
    \item \textbf{Task Execution Tampering Capability:} The adversary can control the malicious node $j$. After accepting the offloaded task $k$, it discards the task with probability $p_{drop}\in[0,1]$, or returns a tampered, incorrect calculation result and forged result hash with probability $p_{tamper}\in[0,1]$ \cite{liang2022cross}.
    \item \textbf{Sybil Attack Capability:} The adversary can try to forge multiple fake node identities to access the network \cite{liu2022bcmaster}. However, it is limited by the admission mechanism of the consortium blockchain, and can only initiate identity forgery through authorized malicious nodes, and cannot access anonymously. Meanwhile, it can collude with multiple controlled malicious nodes to manipulate the reputation feedback of the system.
    \item \textbf{Free-Riding Capability:} The adversary can control rational selfish nodes \cite{guo2026flet}. After accepting the task, it executes the task with the minimum computing power or refuses to execute the task to save its own computing, communication, and flight energy consumption. While it maintains the online state of the node to accept more tasks.
    \item \textbf{Denial of Service (DoS) Attack Capability: } The adversary can control malicious nodes to flood a large number of fake/invalid computing task requests to high-reputation honest service nodes \cite{zhi2025hybrid}. Thus, it can exhaust the computing and communication resources of the target node, resulting in congestion of its task queue and a sharp increase in processing delay.
    \end{itemize}
    
\subsubsection{Adversary Types} Based on the behavior, and attack mode of the adversary, the malicious nodes controlled by the adversary are divided into two categories:
\begin{itemize}
    \item \textbf{Byzantine Malicious Adversary:} Corresponding to the malicious aircraft, its core goal is to actively destroy the security and availability of the system, and launch all the above four types of attacks indiscriminately. 
 \item \textbf{Rational Selfish Adversary:} Corresponding to the free-riding nodes, its core goal is to maximize its own resource income and minimize its own energy consumption. It only initiates selfish free-riding attacks, and will not actively destroy the system.
\end{itemize}

\subsection{Problem Formulation}
Our objective is to find an optimal policy $\pi^*$ that minimizes the long-term, weighted total cost of the system, which is a combination of task completion latency, total energy consumption, and a negative trust-weighted reward. This reflects a core trade-off in the system, which we term the ``Trust-Performance-Overhead" trilemma.

\emph{1) Performance vs. Overhead:} To pursue extremely low latency, an aircraft might prefer to offload tasks to its nearest neighbor. However, recording this transaction on the blockchain for security introduces additional latency and energy overhead, thereby reducing overall performance.

\emph{2) Trust vs. Performance:} The aircraft with the highest reputation may not be the geographically closest or have the most available computational resources. Choosing it might increase latency and energy consumption, but ensures the secure and reliable completion of the task. Conversely, choosing a nearby but low-reputation node is risky but potentially faster.

\emph{3) Trust vs. Overhead:} To maintain a high-precision trust evaluation, more frequent on-chain interactions are needed to update reputation scores, but this directly increases the total overhead of the blockchain system.

Additionally, the optimal decision is not static but highly dependent on the current system state, e.g., task urgency, node reputation, and network congestion. Therefore, our optimization goal is to learn a policy that can dynamically balance these competing factors.

\textbf{Objective Function}:
\begin{equation}
    \begin{aligned}
\label{eq:objective}
\min_{\pi} \lim_{T \to \infty} \frac{1}{T} \sum_{t=1}^{T} \mathbb{E}_{\pi} \bigg [  \sum_{i=1}^{N} \sum_{k} \bigg ( &w_D T_{i,k}^{\text{total}}(t) + w_E E_i^{\text{total}}(t) - \\& w_R \sum_{j \neq i} \mathcal{R}_j(t) a_{ijk}(t) \bigg ) \bigg ]
    \end{aligned}
\end{equation}
where $T_{i,k}^{\text{total}}(t)$ is the end-to-end total latency for task $k$ originating from agent $i$, $E_i^{\text{total}}(t)$ is the total energy consumption of aircraft $i$ at time slot $t$ (including propulsion, computation, communication, and blockchain), $\mathcal{R}_j(t)$ is the reputation of the service provider $j$, and $a_{ijk}(t)$ is the offloading decision indicator. $w_D, w_E, w_R$ are weighting factors to balance the importance of latency, energy, and trust.

\textbf{Constraints}:
\begin{equation}
    \begin{aligned}
    & (C1): \|\mathbf{q}_n(t+1) - \mathbf{q}_n(t)\| \le v_{\max} \Delta t, \quad \forall n \in \mathcal{N}, \\
    & (C2): \|\mathbf{q}_i(t) - \mathbf{q}_j(t)\| \ge d_{\min}, \quad \forall i \neq j \in \mathcal{N}, \\
    & (C3): \sum_{j \in \mathcal{N}} a_{ijk}(t) = 1, \quad \forall i, k, \\
    & (C4): f_n(t) \le f_n^{\max}, \quad P_n^{\text{trans}}(t) \le P_n^{\max}, \quad \forall n \in \mathcal{N}, \\
    & (C5): \Pr\left(T_{i, k}^{\text{total}}(t) \leq T_{k}^{\text{max}}\right) \geq \varepsilon_{k}, \quad \forall i, k.
    \end{aligned}
\end{equation}

\emph{1) Collision Avoidance / Safety Distance Constraint $(C2)$:} The left side of the inequality calculates the distance traveled by aircraft $n$ within a single time slot $\Delta t$. The right side represents the maximum possible distance the aircraft can travel within that time slot at its maximum speed $v_{\max}$. Therefore, this constraint ensures that the aircraft's trajectory is physically feasible, adhering to its inherent speed limitations.

\emph{2) Aircraft Kinematics Constraint $(C1)$:} The left side of the inequality represents the Euclidean distance between any two distinct aircraft, $i$ and $j$. $d_{\min}$ is the predefined minimum safety distance. This constraint is fundamental to the safe operation of the LACNets to prevent mid-air physical collisions.

\emph{3) Task Assignment Uniqueness Constraint $(C3)$:} The summation over all possible computing nodes $j$, including the agent $i$ itself, being equal to $1$ ensures that each computation task is assigned to one and only one location for processing. This prevents task duplication and ensures that no task is dropped during the assignment process.
    
\emph{4) Hardware / Resource Limitation Constraint $(C4)$:} This constraint reflects the physical hardware limitations of each aircraft. The CPU frequency used by aircraft $n$, $f_n(t)$, cannot exceed its processor's maximum clock speed, $f_n^{\max}$. Likewise, its transmission power, $P_n^{\text{trans}}(t)$, is capped by the maximum power supported by its communication module, $P_n^{\max}$. This ensures that the framework's decisions are practical and operate within the feasible capabilities of the agents' hardware.
    
\emph{5) Quality of Service (QoS) Constraint $(C5)$:} $T_{i,k}^{\text{total}}(t)$ represents the total end-to-end latency for completing task $k$. $T_k^{\max}$ is the predefined maximum tolerable deadline for that specific task. We adopt a chance-constrained formulation, where $\varepsilon_k \in (0,1]$ denotes the required completion reliability for task $k$.
This formulation guarantees long-term service quality while tolerating instantaneous violations caused by fading. 

\section{SkyChain Intelligence Solution}\label{sec-Iv}

To solve the complex optimization problem formulated in the previous section, we design the SkyChain Intelligence framework. This framework empowers LAEAI agents with agentic AI capabilities to make autonomous decisions. 

\begin{figure*}[!t]
   \centering
   \includegraphics[width=6 in]{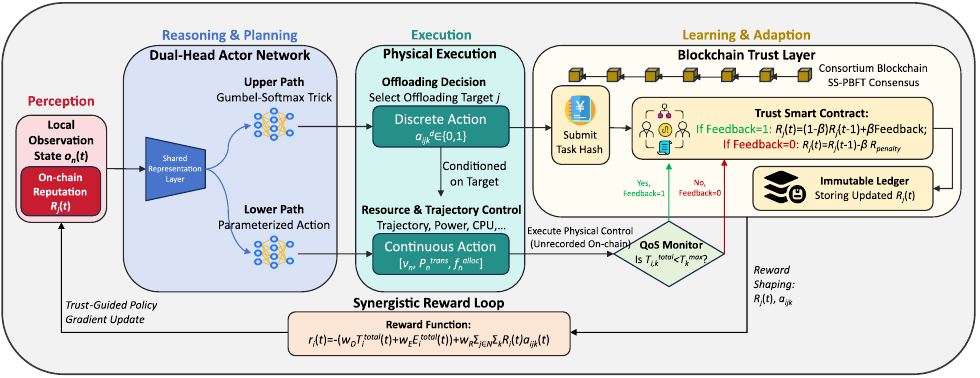}
   \caption{Integrating Agentic Paradigms and Blockchain Trust. It reveals the four LAEAI stages of ``perception-reasoning-execution- adaptation" in the SkyChain framework, deeply coupling the continuous optimization of underlying physical resources with the discrete blockchain reputation reward mechanism.}
   \label{fig_AG}
    \vspace{-0.4cm}
\end{figure*}

\subsection{The Role of Agentic AI}

Agentic AI refers to systems that can autonomously and proactively pursue goals with limited human supervision by perceiving their environment, reasoning, planning, and executing actions \cite{zhang2024generative}, \cite{gao2025agentic}. In our framework, each LAEAI agent is an instance of agentic AI. Its agentic nature is not a pre-programmed set of rules but an emergent property derived from the learning process \cite{zhang2026toward}. The MADRL algorithm is the engine that drives this agentic behavior, as shown in Fig. \ref{fig_AG}.

\emph{1) Perception:} The agent perceives its local environment through its observation space $o_n(t)$, which includes its physical state, network conditions, task queues, and trust information from the blockchain.

\emph{2) Reasoning and Planning:} The actor-critic neural networks process these observations to reason about the complex trade-offs. The agent implicitly plans a sequence of actions, such as trajectory adjustments and offloading decisions, to maximize its long-term cumulative reward.

\emph{3) Execution:} The agent executes its plan by outputting a hybrid action $a_n(t)$ that directly controls its physical movement and resource management decisions.

\emph{4) Learning and Adaptation:} By receiving reward feedback and observing the consequences of its actions, the agent continuously adapts its internal policy, improving its ability to achieve its goals in a dynamic and unpredictable environment.

This approach enables LAEAI agents to move beyond simple automation and exhibit intelligent, goal-directed behavior in the complex low-altitude domain.

\subsection{Agent Decision Process as a Dec-POMDP}

Since each agent has only partial information about the global state and its decisions impact the entire system, we model the problem as a Dec-POMDP defined by the tuple $\langle \mathcal{N}, S, \{A_n\}_{n \in \mathcal{N}}, T, R, \{O_n\}_{n \in \mathcal{N}}, \gamma \rangle$ \cite{kopic2024collaborative}:

\emph{1) State Space ($S$):} The global state $s(t) \in S$ includes all UAVs' positions, velocities, task queues, remaining energy, and the blockchain's current reputation scores.

\emph{2) Observation Space ($O_n$):} At time slot $t$, each agent $n$'s local observation $o_n(t) \in O_n$ is a subset of the global state, including its own status, information about its neighbors, current task properties, and a local copy of the reputation table.

\emph{3) Action Space ($A_n$):} Each agent's action space is hybrid, comprising i) Discrete Action $a_n^d$: The task offloading decision, selecting a computation node $j$ from its neighbors or itself; ii) Continuous Action $a_n^c$: Trajectory control (velocity vector $[v_x, v_y]$), transmit power $P_n^{\text{trans}}$, and CPU frequency $f_n^{\text{alloc}}$.

\emph{4) Transition Probability ($T$):} $T(s'|s, \mathbf{a})$ denotes the probability of transitioning from state $s$ to $s'$ after the joint action $\mathbf{a} = (a_1, \dots, a_N)$.

\emph{5) Reward Function ($R_n$):} Each agent $n$ receives a local reward, as shown below:
    
    \begin{equation}
    \begin{aligned}
        r_i(t) = &-\underbrace{(w_D T_i^{total}(t) + w_E E_i^{total}(t))}_{\text{Basic Performance Cost Term}}
\\&+ \underbrace{w_R \sum_{j \in \mathcal{N}} \sum_k \mathcal{R}_j(t) a_{ijk}(t)}_{\text{Trust Incentive Term}}
,
 \end{aligned}
    \end{equation}

\emph{6) Discount Factor ($\gamma$):} A discount factor $\gamma \in [0, 1)$ that balances immediate and future rewards.


\subsection{Hybrid Action Space MADDPG Algorithm}

We adapt the MADDPG algorithm,  which follows the Centralized Training, Decentralized Execution (CTDE) paradigm, to handle our hybrid action space using a parameterized action space approach.

\emph{1) Actor Network Architecture:} Each agent's actor network $\mu_{\theta_n}$ has two output heads: a \textbf{discrete action head} that uses the Gumbel-Softmax trick to select an offloading target, and a \textbf{continuous parameter head} that outputs the continuous control parameters (trajectory, power, CPU) conditioned on the selected discrete action. The former's decision will be recorded on the blockchain, while the latter will only write the judgment result of $T_{i,k}^{\text{total}}(t) \le T_k^{\max}$ onto the chain. It can avoid the overhead of recording a large number of decisions on the chain. To mitigate the off-chain continuous parameter falsification risk, we design a lightweight hash commitment audit scheme:
 \textit{i) Commitment phase}: The cryptographic hash of negotiated continuous parameters (allocated CPU frequency, transmit power) is recorded on-chain with the discrete offloading decision. Plaintext parameters are transmitted off-chain;
 \textit{ii) Verification phase}: After task completion, the requester infers the actual resource allocation from observed end-to-end latency, and compares its hash with the on-chain commitment;
\textit{iii) Penalty mechanism}: Verified fraud triggers severe reputation deduction and on-chain credit forfeiture.
This scheme ensures full auditability.

\emph{2) Critic Network Architecture:} During training, a centralized critic network $Q_{\phi_n}$ for each agent evaluates the joint action of all agents, taking all observations and actions $(\mathbf{o}, \mathbf{a})$ as input to produce a global Q-value. This guides the actors toward cooperative strategies.

\emph{3) Training Process:} The training follows the standard MADDPG update rules. The critic network $Q_{\phi_n}$ is updated by minimizing the Mean Squared Bellman Error (MSBE) loss function, as shown below:
    \begin{equation}
    \mathcal{L}(\phi_n) = \mathbb{E}_{(\mathbf{o}, \mathbf{a}, \mathbf{r}, \mathbf{o'}) \sim \mathcal{D}} \left[ \left( y_n - Q_{\phi_n}(\mathbf{o}, \mathbf{a}) \right)^2 \right]
    \end{equation}
    where the target value $y_n$ is calculated using the target networks, as shown below:
    \begin{equation}
    y_n = r_n + \gamma Q'_{\phi'_n}(\mathbf{o'}, \mathbf{a'})|_{a'_j = \mu'_{\theta'_j}(o'_j)}
    \end{equation}
    The actor network $\mu_{\theta_n}$ is updated using the sampled policy gradient, as shown below:
    \begin{equation}
        \begin{aligned}
    \nabla_{\theta_n} J(\mu) \approx \mathbb{E}_{\mathbf{o} \sim \mathcal{D}} \big[ &\nabla_{a_n} Q_{\phi_n}(\mathbf{o}, a_1, \dots, a_N)|_{a_n=\mu_{\theta_n}(o_n)} \\&\nabla_{\theta_n} \mu_{\theta_n}(o_n) \big].
        \end{aligned}
    \end{equation}

\begin{algorithm}[!t]
\caption{SkyChain Intelligence Training Algorithm}\label{alg:skychain}
\SetKwInOut{Input}{Input}
\SetKwInOut{Output}{Output}

\underline{Initialization:}\\
\For{each agent $n=1, \dots, N$}{
    1. Initialize actor $\mu_{\theta_n}$ and critic $Q_{\phi_n}$ networks with random weights $\theta_n, \phi_n$ \\
    2. Initialize target networks $\mu'_{\theta'_n}$ and $Q'_{\phi'_n}$ with weights $\theta'_n \leftarrow \theta_n, \phi'_n \leftarrow \phi_n$
}
3. Initialize shared replay buffer $\mathcal{D}$
\BlankLine

\For{episode = 1 to MaxEpisodes}{
    \underline{Episode Initialization:}\\
    1. Receive initial global state $s$ and local observations $\{o_n\}_{n=1}^N$
    
    \For{$t=1$ to MaxSteps}{
        \underline{Action Selection:}\\
        \For{each agent $n=1, \dots, N$}{
            1. Select hybrid action $a_n = (a_n^d, a_n^c)$ from actor $\mu_{\theta_n}(o_n)$ with exploration noise
        }
        
        \underline{Environment Interaction:}\\
        1. Execute joint action $\mathbf{a} = (a_1, \dots, a_N)$ \\
        2. Observe joint reward $\mathbf{r} = (r_1, \dots, r_N)$ and next observations $\{o'_n\}_{n=1}^N$ \\
        3. Store $(\mathbf{o}, \mathbf{a}, \mathbf{r}, \mathbf{o'})$ in replay buffer $\mathcal{D}$ \\
        4. $\mathbf{o} \leftarrow \mathbf{o'}$
        
        \underline{Network Update:}\\
        \For{each agent $n=1, \dots, N$}{
            1. Sample a random minibatch of $S$ transitions $(\mathbf{o}^i, \mathbf{a}^i, \mathbf{r}^i, \mathbf{o'}^i)$ from $\mathcal{D}$ \\
            2. Set target Q-value: $y_n^i = r_n^i + \gamma Q'_{\phi'_n}(\mathbf{o'}^i, \mathbf{a'}^i)|_{a'_j = \mu'_{\theta'_j}(o'_j)}$ \\
            3. Update critic by minimizing the loss: $\mathcal{L}(\phi_n) = \frac{1}{S} \sum_i (y_n^i - Q_{\phi_n}(\mathbf{o}^i, \mathbf{a}^i))^2$ \\
            4. Update actor using the sampled policy gradient: 
            \[
            \begin{split}
            \nabla_{\theta_n} J \approx \frac{1}{S} \sum_i &\nabla_{a_n} Q_{\phi_n}(\mathbf{o}^i, a_1^i, \dots, a_N^i)\\
            &|_{a_n=\mu_{\theta_n}(o_n^i)} \cdot \nabla_{\theta_n} \mu_{\theta_n}(o_n^i)
            \end{split}
            \]
        }
        
        \underline{Target Network Update:}\\
        1. Soft update target networks for each agent $n$: \\
        2. $\phi'_n \leftarrow \tau \phi_n + (1-\tau)\phi'_n$ \\
        3. $\theta'_n \leftarrow \tau \theta_n + (1-\tau)\theta'_n$
    }
}
\end{algorithm}

    The overall process is detailed in Alg. \ref{alg:skychain}.

\subsection{Computational Complexity Analysis}

The computational complexity of the SkyChain Intelligence solution is analyzed by separating it into two distinct phases: 

\emph{1) Decentralized Execution Complexity:} This is the most critical aspect for real-time LAEAI operations. During execution, each agent $n$ only needs to perform a single forward pass through its local actor network $\mu_{\theta_n}$ to compute its action $a_n$ from its local observation $o_n$. Let $L_a$ represent the number of parameters in the actor network's layers. The computational complexity for this operation is $\mathcal{O}(L_a)$. Crucially, this complexity is independent of the total number of agents $N$ in the network. This low and constant-time complexity ensures that the decision-making process is highly scalable and suitable for deployment on resource-constrained UAVs.

\emph{2) Centralized Training Complexity:} The training phase, typically performed offline on a powerful server, bears the majority of the computational load. The complexity is driven by the centralized critic updates. Each of the $N$ critic networks, $Q_{\phi_n}$, must process the joint observations $\mathbf{o} = (o_1, \dots, o_N)$ and joint actions $\mathbf{a} = (a_1, \dots, a_N)$ of all $N$ agents. Consequently, the input size of each critic scales linearly with $N$. Let $S$ be the minibatch size, and $L_c$ be the complexity of a critic network. To update all $N$ critics and $N$ actors for one gradient step, the total computational complexity is approximately $\mathcal{O}(S \cdot N^2 \cdot (L_a + L_c))$. This quadratic scaling $\mathcal{O}(N^2)$ with the number of agents is a known characteristic of MADDPG, representing the trade-off for achieving globally-aware, cooperative policies.


In summary, the SkyChain Intelligence framework intelligently shifts the high computational burden, quadratic scaling $\mathcal{O}(N^2)$ to the offline training phase, while maintaining a low, constant-time complexity, $\mathcal{O}(1)$ with respect to $N$, for real-time, decentralized execution.

\section{Security Analysis}\label{sec-v}
The SkyChain Intelligence framework is designed to be resilient against various security threats prevalent in decentralized multi-agent systems. This section analyzes its robustness against several common attack vectors.
\subsection{Data Tampering and Integrity Attacks}

This attack represents a malicious service-providing agent that can execute a task incorrectly, either intentionally or due to a fault, and return a false result to the requester to save computational resources \cite{jia2024parameter}.

Our framework counters this threat using the blockchain's inherent properties of immutability and transparency. When a task is completed, the service provider submits the hash of the computation result, which is recorded in an immutable transaction on the blockchain. The task requester can verify the result and submit feedback, success or failure, to the reputation smart contract. Any attempt by the provider to tamper with the result would produce a different hash, which the requester can easily detect. Consistent submission of incorrect results will lead to negative feedback, causing the malicious agent's reputation score to plummet. The DRL agents, driven by the reputation-fused reward function, will learn to avoid offloading tasks to this untrustworthy node, effectively isolating it from the network's economy.
\subsection{Sybil Attacks}
  It is an adversary that can create a large number of pseudonymous identities, namely Sybil nodes, to gain a disproportionate influence on the network, manipulate the reputation system, or launch coordinated attacks. 

The SkyChain framework employs a two-layered defense. First, the use of a consortium blockchain acts as a primary barrier. Unlike public blockchains, nodes cannot join anonymously. They must be authenticated and granted permission to participate. This makes creating a large number of fake identities significantly more difficult and costly for an attacker. Second, even if an attacker manages to introduce a few malicious nodes, the reputation system provides further resilience. New nodes start with a neutral or low reputation and must build trust over time through reliable interactions. The cost and time required to build a positive reputation for a multitude of Sybil nodes make it economically infeasible to manipulate the system on a large scale.

\subsection{Selfish Behavior and Free-Riding}

Rational but selfish agents might agree to provide a service but then fail to execute the task or execute it with minimal resources, to conserve their energy and computility, a behavior known as free-riding \cite{chen2024toward}.

 The synergistic link between the blockchain's reputation score and the MADRL's reward function directly disincentivizes such behavior. An agent that consistently fails to complete tasks will receive negative feedback, leading to a low reputation score. Since other agentic AIs are optimizing for a reward that includes the reputation of their partners, they will quickly learn that selecting a selfish, low-reputation node results in a lower cumulative reward. Consequently, selfish agents will receive fewer and fewer offloading requests, denying them the potential benefits of participating in the computational market and isolating them from the cooperative network.

 \subsection{Denial of Service (DoS) Attacks}
 
An attacker could attempt to disrupt the network by flooding a high-reputation, critical service-providing aircraft with a large volume of malicious or fake task requests, aiming to overwhelm its resources and render it unavailable to legitimate users \cite{uddin2024denial}.  

While the blockchain layer does not directly prevent DoS attacks, the agentic AI layer provides significant resilience through adaptive decision-making. Each agent's observation space includes local metrics like task queue length and experienced latency. When a node is under a DoS attack, its processing and queuing delays will increase dramatically. Neighboring agents will observe this degradation in performance. Their MADRL policies, trained to minimize overall task completion latency, will naturally learn to avoid the congested node and reroute their offloading requests to other available, responsive agents. This intelligent, decentralized load-balancing behavior effectively mitigates the impact of the DoS attack on the overall network performance without requiring a central authority to detect and block the attack.

\section{Performance Evaluation}\label{sec-vi}
This section provides a comprehensive evaluation of the proposed SkyChain Intelligence framework through extensive simulation experiments using a MATLAB-NS-3 co-simulation architecture, where NS-3 emulates realistic wireless propagation, UAV mobility and blockchain consensus interactions, and MATLAB implements the MADRL algorithm and system performance calculation. It runs on a server equipped with three 96-core Intel(R) Xeon(R) Gold 5220R CPUs, 1 TB of memory,
and 8 NVIDIA GeForce RTX 3090 GPUs. 

\subsection{Simulation Setup and Comparison Schemes}
We simulate a 3D environment over a $2000$ m x $2000$ m 2D area, containing $50$ randomly distributed ground task generators. The simulation proceeds in discrete time slots, running for a total of $2000$ slots. All simulation results are averaged over $50$ independent experiments to ensure statistical validity. The other key parameters used in the simulation are detailed in Table \ref{tab:sim_params}. 

\begin{table}[t!]
\centering
\caption{Simulation Parameters}
\label{tab:sim_params}
{\footnotesize 
\begin{tabularx}{8.8cm}{ 
  >{\hsize=0.55\hsize\raggedright\arraybackslash}X  
  >{\hsize=0.45\hsize\raggedright\arraybackslash}X   
}
\hline\hline
\textbf{Parameter} & \textbf{Value} \\
\hline
\multicolumn{2}{c}{\textbf{Network Settings}} \\
\hline
Area Size & $2000 \times 2000$ m$^2$ \\
Number of Aircraft & 5, 10, 15, 20, 25 \\
Percentage of Malicious aircraft & 10\% \\
Number of Task Generators & 50 \\
Aircraft Flight Altitude ($H$) & Uniformly in [50, 100] m \\
Simulation Time Slots & 2000 \\
\hline
\multicolumn{2}{c}{\textbf{Communication Parameters}} \\
\hline
Channel Model & Probabilistic LoS (Urban) \\
Path Loss Parameters ($a, b$) & 9.61, 0.16 \\
Bandwidth ($B$) & 20 MHz \\
Transmit Power ($P_m$) & 0.1 W \\
Noise Power Spectral Density ($N_0$) & -174 dBm/Hz \\
Shadowing Std. (LoS / NLoS) & 4 dB / 12 dB \\
Rician K-factor (A2A LoS) & 20 dB \\
\hline
\multicolumn{2}{c}{\textbf{Computation Task Parameters}} \\
\hline
Task Data Size ($D_k$) & Uniformly in [0.5, 2.0] Mbits \\
Task Computation Density ($C_k$) & Uniformly in [500, 2500] cycles/bit \\
Task Arrival Rate ($\lambda$) & Uniformly in [0.1, 0.9] tasks/slot/device \\
Aircraft CPU Frequency ($f_n$) & Uniformly in [1.0, 2.0] GHz \\
\hline
\multicolumn{2}{c}{\textbf{Ariel Model Parameters}} \\
\hline
Max Speed ($v_{\max}$) & 25 m/s \\
Propulsion Power Params ($P_0, P_i$) & 79.8 W, 88.6 W \\
Min Safety Distance ($d_{\min}$) & 50 m \\
Completion Reliability ($\varepsilon_k$) & 0.95\\
\hline
\multicolumn{2}{c}{\textbf{Blockchain Parameters}} \\
\hline
Block Interval & 2 s \\
Transaction Size & 256 Bytes \\
Consensus Message Size ($S_{\text{msg}}$) & 128 Bytes \\
Minimum Consensus Rate ($R_{\text{min}}$) & 1 Mbps \\
Local Validation Energy ($E_{\text{comp}}^{\text{BC}}$) & 0.002 J \\
Shard LoS Threshold ($P_{\text{th}}^{\text{shard}}$) & 0.8 \\
Short-term Reputation Factor ($\beta_s$) & 0.3 \\
Long-term Reputation Factor ($\beta_l$) & 0.05 \\
Reputation Fusion Weight ($\lambda$) & 0.4 \\
Checkpoint interval ($K$) & 10 blocks \\
\hline
\multicolumn{2}{c}{\textbf{MADDPG Parameters}} \\
\hline
Actor/Critic Network & 3-layer MLP (256-256-128) \\
Learning Rate (Actor/Critic) & $1 \times 10^{-4}$ / $1 \times 10^{-3}$ \\
Discount Factor ($\gamma$) & 0.99 \\
Replay Buffer Size & $1 \times 10^6$ \\
Batch Size & 1024 \\
Objective Weights ($w_D, w_E, w_R$) & (0.4, 0.4, 0.2) \\
\hline\hline
\end{tabularx}
}
 \vspace{-0.4cm}
\end{table}

To validate the superiority of our proposed SkyChain framework, we compare it with the following three comparison schemes:
1) Game-Theoretic Offloading (GTO) \cite{chen2024multi} is a non-cooperative game-theoretic approach in which each LAEAI agent acts selfishly to maximize its utility; 
2) FL-Aided Offloading (FLAO) \cite{tummala2024efficient} is a scheme where agents use FL to collaboratively train a shared neural network model that predicts the expected latency and energy cost for offloading a task to any other agent in the network; 
3) DRL without Blockchain (DRL-noBC) \cite{tang2025task} uses a similar MADDPG algorithm without the blockchain layer and trust mechanism, but retains the core of Agentic AI. 
4) Blockchain-Empowered MADRL (BC-MADRL) \cite{jia2025trusted} is a homologous blockchain-aided multi-agent reinforcement learning baseline adapted for computation offloading. It leverages a trust-enhanced PBFT consensus protocol to maintain dynamic node reputation scores.

\subsection{Convergence comparison}
\begin{figure*}[!t]
   \centering
   \includegraphics[width=7 in]{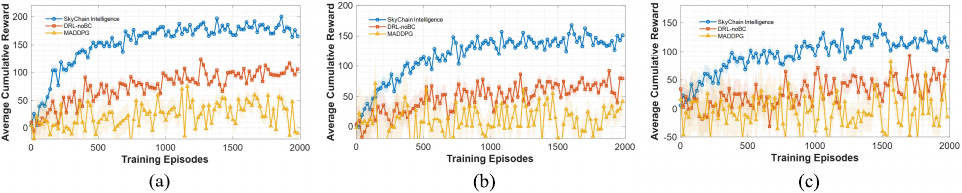}
   \caption{Convergence comparison. (a) Baseline normal scenario; (b) High network load scenario; (c). High security threat scenario.}
   \label{figcon}
    \vspace{-0.4cm}
\end{figure*}

We verify the convergence of SkyChain Intelligence by comparing it with DRL-noBC and MADDPG from \cite{chen2025maddpg} in three scenarios. The three scenarios are as follows. 1) Baseline normal scenario: $20$ aircraft, $10\%$ malicious nodes, task arrival rate of $\lambda=0.5$ tasks/slot/device; 2) High network load scenario: $20$ aircraft, $10\%$ malicious nodes, task arrival rate $\lambda=0.9$ tasks/slot/device; 3) High security threat scenario: $20$ aircraft, $20\%$ malicious nodes, task delivery rate $\lambda=0.5$ tasks/slot/device.  The result is shown in Fig. \ref{figcon}.

As shown in Fig. \ref{figcon} (a), in the baseline conventional scenario, SkyChain Intelligence has the fastest convergence speed, reaching $95\%$ stable rewards after $300$ rounds. Its training process shows no significant fluctuations. The convergence speed of DRL-noBC is $60\%$ slower than that of SkyChain Intelligence, and the stable reward is only $55.6\%$ of that of SkyChain Intelligence. MADDPG has even worse convergence performance, with a mean stable reward of only $40$ and a highly fluctuating training process.
As shown in Fig. \ref{figcon} (b), in the high network load scenario, SkyChain Intelligence still maintains stable convergence. After $350$ rounds, it reaches $95\%$ steady-state rewards, with a small fluctuation range that is controllable. The convergence speed and performance of DRL-noBC deteriorate significantly. After $650$ rounds, it reaches the convergence threshold, which is $30\%$ lower than the benchmark scenario. MADDPG completely stagnates in convergence, and no effective strategy learning occurred within $2000$ rounds.
As shown in Fig. \ref{figcon} (c), in the high security threat scenario, the convergence performance of SkyChain Intelligence slightly deteriorated but remained stable. It reaches $95\%$ steady-state reward after $400$ rounds. Compared to the baseline scenario, it only decreases by $33.3\%$, achieving precise isolation of malicious nodes. The performance of DRL-noBC plummets. It takes 880 rounds to reach the convergence threshold. MADDPG failed to converge, deteriorating by $87.5\%$ compares to the baseline scenario and failing strategy learning.

The reason SkyChain Intelligence has an advantage in convergence is as follows: First, dynamic reputation scoring on the blockchain enables the active isolation of malicious nodes. This eliminates the high noise and strong randomness of the reward signals caused by task failures and retransmissions; Secondly, embodied AI allows agents to autonomously adapt to the high dynamics and non-stationary characteristics of low-altitude networks, avoiding the strategy failure and convergence oscillation caused by environmental dynamic changes; Thirdly, the dual-head actor-critic architecture breaks through the action space limitations of the MADDPG. It can achieve end-to-end global optimization of discrete offloading decisions and continuous resource allocation. Also, it directly embeds the on-chain reputation score into the reward function, which can simultaneously improve the convergence speed.

\subsection{Performance Comparison}


  We present and analyze the simulation results under the $10\%$ of the aircraft are malicious. These malicious agents accept offloading tasks but have a high probability ($90\%$) of dropping them to save their resources, effectively failing the task. This scenario is designed to test the robustness and practical value of our framework in the open wireless LACNets.

\begin{figure*}[!t]
   \centering
   \includegraphics[width=7 in]{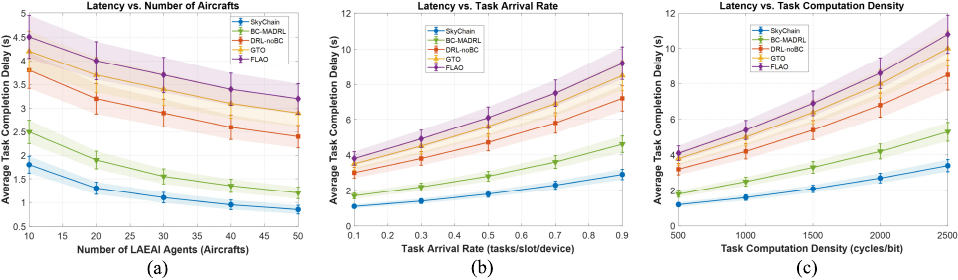}
   \caption{Average task completion delay. (a) Under different numbers of LAEAI agents; (b) Under different task arrival rates; (c) Under different task computation densities.}
   \label{fig3}
    \vspace{-0.4cm}
\end{figure*}

  \textbf{\emph{1) Average Task Completion Delay:}}  As shown in Fig. \ref{fig3} (a), as the number of aircraft increases, the average latency for SkyChain decreases significantly, as it learns to route tasks to the growing pool of reliable nodes. BC-MADRL, as a blockchain-aided baseline, also outperforms all trust-agnostic schemes, but its latency remains consistently higher because it only incorporates trust into the observation space without reward-level deep fusion. In stark contrast, the latency for DRL-noBC, GTO, and FLAO remains high and decreases much more slowly. These trust-agnostic schemes frequently offload tasks to malicious nodes, leading to task failures and necessitating re-transmissions, which severely inflates their average latency. SkyChain's ability to identify and avoid malicious nodes gives it a decisive performance advantage.
As shown in Fig. \ref{fig3} (b), with increasing network load, the latency of all schemes rises. However, the performance gap between SkyChain and the other schemes widens dramatically. BC-MADRL sees moderate latency growth, but under high load, its insufficient trust incentive leads to more misselections of low-reputation nodes, enlarging the gap with our proposal. For DRL-noBC, GTO, and FLAO, a higher load means that more tasks are inevitably sent to malicious nodes, resulting in a catastrophic rise in effective latency. SkyChain, however, consistently routes tasks to trusted nodes, and its latency increases gracefully due to normal network congestion, showcasing its superior robustness under pressure.
As shown in Fig. \ref{fig3} (c), as tasks become more computationally intensive, the cost of a failed offload becomes much higher. Consequently, the latency of DRL-noBC, GTO, and FLAO skyrockets. BC-MADRL mitigates most malicious failures via its basic reputation mechanism, but lacks hybrid-action joint optimization for precise resource matching, so its latency grows faster than SkyChain. SkyChain, by ensuring tasks are sent to reliable nodes from the outset, avoids the crippling penalty of failed complex tasks, demonstrating a clear and widening performance gap as task complexity increases.

\begin{figure*}[!t]
   \centering
   \includegraphics[width=7 in]{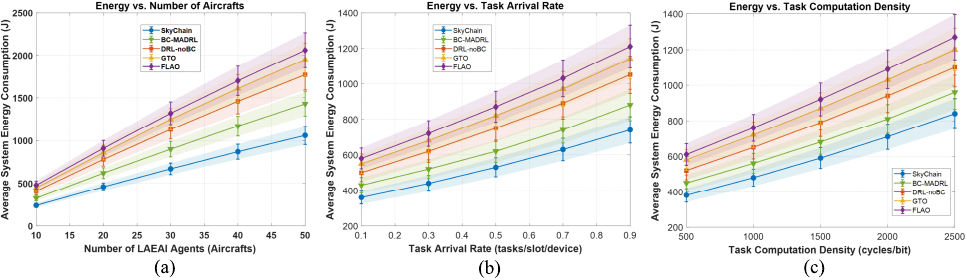}
   \caption{Average system energy consumption. (a) Under different numbers of LAEAI agents; (b) Under different task arrival rates; (c) Under different task computation densities.}
   \label{fig4}
    \vspace{-0.4cm}
\end{figure*}

  \textbf{\emph{2) Average System Energy Consumption:}} In Fig. \ref{fig4} (a), it demonstrates that the total system energy for SkyChain has a moderate, linear increase with more aircraft. BC-MADRL outperforms trust-agnostic schemes but consumes more energy than SkyChain, as its basic trust mechanism only partially eliminates retransmission waste. The energy consumption for DRL-noBC, GTO, and FLAO is significantly higher. This is because every task offloaded to a malicious node wastes the transmission energy and requires re-transmission, doubling the energy cost for that task. SkyChain avoids this wasted energy by making reliable offloading decisions, thus achieving much higher energy efficiency for the entire system.
As the task rate increases, the energy wastage of the trust-agnostic schemes is amplified, which is illustrated in Fig. \ref{fig4} (b).  BC-MADRL sees a moderate growth rate, but its shallow trust integration leads to accumulating retransmission overhead under high load. More tasks being sent to malicious nodes leads to a rapid increase in total energy consumption for DRL-noBC, GTO, and FLAO. SkyChain's energy consumption increases at a much slower rate, as it primarily expends energy on productive computations, not on failed and repeated transmissions.
The energy difference is less pronounced with task density, as computation energy at the malicious node is not expended by the requester, as shown in Fig. \ref{fig4} (c). BC-MADRL narrows the gap with SkyChain, but residual retransmission waste still keeps its energy consumption higher. However, the energy wasted on transmission and re-transmission still keeps SkyChain as the most energy-efficient solution, especially when considering the energy cost of not completing high-value tasks.

\begin{figure*}[!t]
   \centering
   \includegraphics[width=7 in]{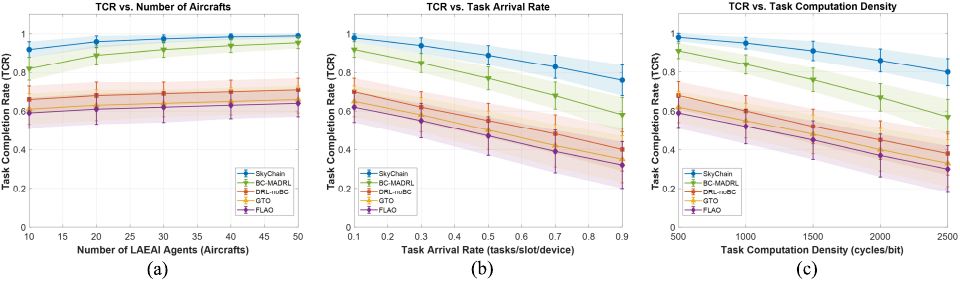}
   \caption{Task completion rate. (a) Under different numbers of LAEAI agents; (b) Under different task arrival rates; (c) Under different task computation densities.}
   \label{fig5}
    \vspace{-0.4cm}
\end{figure*}

  \textbf{\emph{3) Task Completion Rate:}} As shown in Fig. \ref{fig5} (a), this metric most clearly demonstrates the value of SkyChain.this metric most clearly demonstrates the value of SkyChain. The TCR for SkyChain remains consistently high (above \(90\%\)) and improves with more UAVs. BC-MADRL achieves a moderate completion rate, as its basic reputation mechanism filters most malicious nodes, but still lags behind SkyChain due to the lack of trust-performance joint optimization. Conversely, the TCR for DRL-noBC, GTO, and FLAO hovers at a much lower level. Since \(10\%\) of the nodes are malicious and these schemes choose offloading targets without regard to trust, they are statistically prone to a high rate of failure, which our simulation confirms.
 As network load increases, the TCR for the trust-agnostic schemes plummets, as demonstrated in Fig. \ref{fig5} (b). BC-MADRL sees moderate degradation, but its shallow trust integration cannot fully avoid mis-selection of low-reputation nodes under heavy load. Malicious nodes continue to drop tasks, and network congestion causes even legitimate tasks to time out. SkyChain, by efficiently using only the reliable portion of the network, maintains a much more stable and higher TCR, proving its reliability in a hostile environment.
Similarly, as shown in Fig. \ref{fig5} (a), as task complexity rises, the trust-agnostic schemes fail more often due to both malicious activity and their inability to manage resources for demanding tasks. BC-MADRL mitigates partial malicious failures but lacks precise resource matching via hybrid action spaces, leading to more timeouts for computation-intensive tasks. SkyChain's ability to ensure that complex tasks are sent to reliable and capable nodes allows it to maintain a significantly higher completion rate, highlighting its suitability for mission-critical applications.

    \begin{figure*}[!t]
   \centering
   \includegraphics[width=7 in]{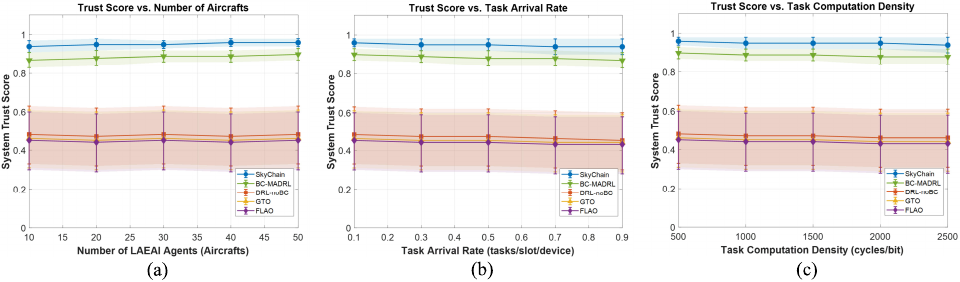}
   \caption{System trust score.  (a) Under different numbers of LAEAI agents; (b) Under different task arrival rates; (c) Under different task computation densities.}
   \label{fig6}
    \vspace{-0.4cm}
\end{figure*}

  \textbf{\emph{4) System Trust Score:}} As shown in Fig. \ref{fig6} (a), the System Trust Score for SkyChain is consistently high (around \(0.95\)), demonstrating that its agents have successfully learned to select high-reputation partners. BC-MADRL maintains a moderately high trust level (around \(0.88\)), as its basic reputation mechanism also guides agents to prefer trustworthy nodes, but the absence of reward-level trust incentive prevents it from achieving optimal trust selection. For DRL-noBC, GTO, and FLAO, the trust score is consistently low (around \(0.45-0.5\)). This is because they have no mechanism to differentiate between trustworthy and malicious nodes, and their offloading decisions are essentially random with respect to trust. In an environment with \(10\%\) malicious nodes, their choices often fall on untrustworthy agents.
As shown in Fig. \ref{fig6} (b), the trust score for all schemes remains relatively stable regardless of the task arrival rate. SkyChain maintains its high trust score, showing that its agents do not compromise on security even when the network is busy. BC-MADRL also retains its moderate trust level without obvious degradation under heavy load. The other schemes remain at their low trust score, continuously making poor, trust-agnostic decisions.
As with the other factors, the trust score remains constant. As shown in Fig. \ref{fig6} (c), the results are stark: SkyChain builds and maintains a highly trustworthy interaction environment, BC-MADRL delivers limited trust improvement via basic blockchain mechanisms, while the other schemes operate in a low-trust environment, making them vulnerable and unreliable. This clearly validates the effectiveness and necessity of deeply fusing the blockchain-based reputation system into the DRL learning framework, rather than simply combining the two technologies.

\subsection{Ablation Experiment}

To verify the necessity of the multiple components for SkyChain, we conduct ablation experiments. They are respectively represented as w/o blockchain, w/o trust reward, w/o Agentic AI, and w/o hybrid action, signifying the absence of blockchain, trust rewards, the perception-reasoning-planning-execution-adaptive closed loop of embodied AI, and the hybrid action space. The ablation test scenarios are consistent with the scenarios used in the convergence comparison.

This set of ablation experiments validates the irreplaceability of each core component of SkyChain Intelligence. In all scenarios, the full SkyChain model achieves optimal performance across all four metrics. As shown in Table \ref{tab:ablation-baseline}, in the baseline normal scenario, it delivers an average delay of $1.6\pm0.2$ s, average energy consumption of $490±25$ J, TCR of $94.1\%\pm1.2\%$, and system trust score of $0.95\pm0.02$. Removing any single component breaks the ``Trust-Performance-Overhead" trilemma balance, with performance degradation following a clear severity pattern: removing the blockchain layer causes the most catastrophic collapse, e.g., TCR drops to $38.2\%\pm3.5\%$ and delay surges to $7.1\pm0.7$ s in the high security threat scenario, followed by removing trust-reward fusion, hybrid action space optimization, and Agentic AI. Notably, environmental complexity amplifies degradation. As shown in Table \ref{tab:ablation-highload} and \ref{tab:ablation-highthreat}, in the high network load and high security threat scenarios, all variants show significantly worse performance, while the full SkyChain model remains robust, with only minor fluctuations. For instance, TCR stays above $88\%$ and trust score above $0.92$ in the high threat scenario.
\begin{table*}[t]
  \centering
  \caption{Results of the Ablation Experiment in Baseline Normal Scenario}
  \label{tab:ablation-baseline}
  \begin{tabular}{lcccc}
    \hline
   Models  & Aver. delay(s) & Aver. energy consumption (J) & Task completion rate & System trust score \\
    \hline
    SkyChain   & $1.6 \pm 0.2$ & $490 \pm 25$         & $94.1\% \pm 1.2\%$ & $0.95 \pm 0.02$ \\
    w/o Trust Reward      & $2.3 \pm 0.3$ & $650 \pm 35$         & $83.5\% \pm 1.8\%$ & $0.74 \pm 0.04$ \\
    w/o Hybrid Action     & $2.9 \pm 0.3$ & $720 \pm 30$         & $78.4\% \pm 2.0\%$ & $0.70 \pm 0.05$ \\
    w/o Agentic AI        & $3.4 \pm 0.4$ & $810 \pm 40$         & $72.6\% \pm 2.2\%$ & $0.65 \pm 0.05$ \\
    w/o Blockchain        & $5.2 \pm 0.5$ & $1080 \pm 45$        & $54.2\% \pm 2.8\%$ & $0.47 \pm 0.03$ \\
    
    \hline
  \end{tabular}
   \vspace{-0.2cm}
\end{table*}

\begin{table*}[t]
  \centering
  \caption{Results of the Ablation Experiment in High Network Load Scenario}
  \label{tab:ablation-highload}
  \begin{tabular}{lcccc}
    \hline
   Models  & Aver. delay(s) & Aver. energy consumption (J) & Task completion rate & System trust score \\
    \hline
    SkyChain   & $2.0 \pm 0.2$ & $550 \pm 30$         & $90.5\% \pm 1.6\%$ & $0.93 \pm 0.03$ \\
    w/o Trust Reward      & $3.0 \pm 0.3$ & $740 \pm 40$         & $78.2\% \pm 2.2\%$ & $0.70 \pm 0.05$ \\
    w/o Hybrid Action     & $3.7 \pm 0.4$ & $820 \pm 35$         & $72.1\% \pm 2.4\%$ & $0.66 \pm 0.06$ \\
    w/o Agentic AI        & $4.5 \pm 0.5$ & $950 \pm 45$         & $63.8\% \pm 2.7\%$ & $0.60 \pm 0.06$ \\
    w/o Blockchain        & $6.3 \pm 0.6$ & $1210 \pm 55$        & $45.7\% \pm 3.3\%$ & $0.44 \pm 0.04$ \\
    
    \hline
  \end{tabular}
   \vspace{-0.2cm}
\end{table*}

\begin{table*}[t]
  \centering
  \caption{Results of the Ablation Experiment in High Security Threat Scenario}
  \label{tab:ablation-highthreat}
  \begin{tabular}{lcccc}
    \hline
   Models  & Aver. delay(s) & Aver. energy consumption (J) & Task completion rate & System trust score \\
    \hline
    SkyChain   & $2.2 \pm 0.3$ & $580 \pm 35$         & $88.7\% \pm 1.8\%$ & $0.92 \pm 0.03$ \\
    w/o Trust Reward      & $3.6 \pm 0.4$ & $830 \pm 45$         & $71.4\% \pm 2.5\%$ & $0.65 \pm 0.05$ \\
    w/o Hybrid Action     & $4.1 \pm 0.4$ & $890 \pm 40$         & $67.5\% \pm 2.6\%$ & $0.63 \pm 0.06$ \\
    w/o Agentic AI        & $4.9 \pm 0.5$ & $1020 \pm 50$        & $59.2\% \pm 2.9\%$ & $0.57 \pm 0.06$ \\
    w/o Blockchain        & $7.1 \pm 0.7$ & $1380 \pm 60$        & $38.2\% \pm 3.5\%$ & $0.41 \pm 0.04$ \\
    
    \hline
  \end{tabular}
   \vspace{-0.2cm}
\end{table*}


\section{Conclusion}\label{sec-vii}
This paper has investigated the core challenges of enabling autonomous, secure, and efficient collaborative LAEAI agents in decentralized LACNets. We have identified and formalized a fundamental ``Trust-Performance-Overhead" trilemma that plagues existing solutions, and proposed SkyChain Intelligence, an innovative framework to address this issue. 
Through comprehensive simulation evaluations, convergence comparison, and ablation experiments, we have validated the superiority and robustness of the proposed SkyChain Intelligence. It has consistently achieved lower end-to-end latency, higher energy efficiency, and greater TSR reliability than advanced baseline schemes. Meanwhile, our framework has maintained fast and stable convergence across complex environments, and ablation studies have verified the indispensable role of each core component. This work has demonstrated that deeply integrating verifiable trust mechanisms into the learning core of agentic AI can enable decentralized autonomous systems that are both efficient and trustworthy. 


\bibliographystyle{IEEEtran}
\bibliography{IEEEabrv,mylib}

\vfill

\end{document}